\documentclass{PoS}

\usepackage{latexsym}
\usepackage{graphicx}
\usepackage{epstopdf}
\usepackage{bm}
\usepackage{color}
\usepackage{fancyhdr}

\title{Finite-size scaling in nucleon axial charge from 2+1-flavor DWF lattice QCD}

\ShortTitle{Finite-size scaling in nucleon axial charge}

\author{Meifeng Lin\thanks{
Current affiliation: Center for Computational Science, Boston University, 3 Cummington Street, Boston, MA 02215, USA.}\\
     Department of Physics, Yale University, New Haven CT 06520, USA\\
	RIKEN BNL Research Center, Brookhaven National Laboratory, Upton, NY 11973, USA\\
     E-mail: \email{meifeng.lin@yale.edu}}
\author{
	\speaker{Shigemi Ohta}\thanks{
		We thank RBC and UKQCD Collaborations, especially Yasumichi Aoki, Tom Blum, Chris 	Dawson, Taku Izubuchi, Chulwoo Jung, Shoichi Sasaki and Takeshi Yamazaki.
RIKEN, BNL, the U.S.\ DOE, University of Edinburgh, and the U.K.\  PPARC provided  facilities essential for the completion of this work.
The I+DSDR ensembles are being generated at ANL Leadership Class Facility (ALCF.)
The nucleon two- and three-point correlators are being calculated at RIKEN Integrated Cluster of Clusters (RICC) and US NSF Teragrid/XSEDE clusters.}\\
	Institute of Particle and Nuclear Studies, KEK, Tsukuba, Ibaraki 305-0801, Japan\\
	Department of Particle and Nuclear Physics, SOKENDAI, Hayama, Kanagawa 240-0193, Japan\\
	RIKEN BNL Research Center, Brookhaven National Laboratory, Upton, NY 11973, USA\\
	E-mail: \email{shigemi.ohta@kek.jp}
}
\author{for RBC and UKQCD Collaborations}

\abstract{
We report the current status of the on-going lattice-QCD calculations of nucleon isovector axial charge, \(g_A\), using the RBC/UKQCD 2+1-flavor dynamical domain-wall fermion ensembles at lattice cutoff of about \(a^{-1}=1.4\) GeV in a spatial volume \((L = 4.6 {\rm fm})^3\).
The result from the ensemble with \(m_\pi = 250\) MeV pion mass, corresponding to the finite-size scaling parameter \(m_\pi L \sim 5.8\), agrees well with an earlier result at \(a^{-1}=1.7\) GeV, \(L = 2.8\) fm, and \(m_\pi = 420\) MeV, with similar \(m_\pi L\).
This suggests the systematic error from excited-state contamination is small in both ensembles and about 10-\% deficit in \(g_A\) we are observing is likely a finite-size effect that scales with \(m_\pi L\). We also report the result from the lighter, \(m_\pi = 170\) MeV ensemble.

\vspace{-174mm}\parbox{\textwidth}{\flushright\large\rm \hfill KEK-TH-1594, RBRC-977}\vspace{171mm}
}

\FullConference{The 30 International Symposium on Lattice Field Theory - Lattice 2012,\\
		June 24-29, 2012\\
		Cairns, Australia}

\begin{document}

\pdfinfo{
  /Title   (Lattice2012)
  /Author  (Shigemi Ohta)
  /Subject (Nucleon Structure)
}

\section{Introduction}

RBC and UKQCD collaborations have been jointly calculating nucleon-structure observables such as form factors and low moments of structure functions \cite{Yamazaki:2008py,Yamazaki:2009zq,Aoki:2010xg,Ohta:2010sr,Lin:2011vx,Ohta:2011vv} using their 2+1-flavor dynamical Domain-Wall Fermions (DWF)  lattice-QCD ensembles \cite{Allton:2008pn,Aoki:2010dy,Arthur:2012yc}.
Here we report the ongoing calculation using the latest ensembles with pion mass, \(m_\pi\), as low as  250 and 170 MeV with a linear spatial extent, \(L\), of about 4.6 fm \cite{Arthur:2012yc}, corresponding to the finite-size scaling parameter, \(m_\pi L\), of about 6 and 4 respectively.

Using our earlier ensembles with pion mass down to about 330 MeV in a linear spatial extent of about 2.8 fm, we observed a significant deficit in the isovector axial charge, \(g_A\), of nucleon, of about 10\% compared with the experiment \cite{Yamazaki:2008py}.
By comparing with many other similar lattice numerical calculations with relatively heavier mass and smaller spatial extents that showed similar trends, we conjectured the deficit is likely a finite-size effect from relatively small values of the scaling parameter, \(m_\pi L\) \cite{Yamazaki:2008py}.
This can be interpreted as the first ever concrete evidence for the ``virtual pion cloud'' surrounding the nucleon: Such a cloud carries isovector axialvector current of the nucleon, and when the pion is light, can fluctuate far away from the nucleon center of mass resulting in a deficit if the volume is insufficient.
However some questions against this interpretation were raised in the past couple of years, that the deficit we and many others had seen might have been systematic errors from excited-state contamination \cite{Wittig:2012np,Capitani:2012gj}.

The preliminary results of our ongoing calculations, with different gauge action, different lattice cut off, much lighter pion mass and much larger lattice spatial extent are already sufficient to exclude such excited-state contamination from at least two of our calculations: the earlier one at \(m_\pi = 420\) MeV and \(L = 2.8\) fm and the ongoing one at \(m_\pi = 250\) MeV and \(L = 4.6\) fm that share about the same value of \(m_\pi L = 5.8\), and reinforce our conjecture of scaling with \(m_\pi L\).

Details of form factors are reported separately by Meifeng Lin  in these proceedings \cite{MeifengLattice2012}.

\section{Numerics}

The joint RBC and UKQCD 2+1-flavor dynamical DWF ensembles \cite{Arthur:2012yc} we are using in this report are generated with a new gauge action, a combination of conventional Iwasaki rectangular-improved action \cite{Iwasaki:1983ck} at the inverse-squared gauge coupling of  \(\beta=1.75\) and a multiplicative disolocation-suppressing determinant ratio (DSDR) factor \cite{Vranas:1999rz,Vranas:2006zk,Renfrew:2009wu}.
This combination allows us calculations with relatively low lattice cut-off momentum of \(a^{-1}\sim 1.371(8)\) GeV without driving up the residual chiral symmetry breaking so the DWF residual mass is \(m_{\rm res}a \sim 0.002\) while maintaining reasonable distribution in gauge field homotopy.
Consequently the linear spatial extent of the lattice, with 32 lattice sites in one direction, is about 4.6 fm, and allows us to calculate at much closer to physical pion mass, \(m_\pi\sim 170\) and 250 MeV, or input quark mass, \(m_{ud}a = 0.001\) and 0.0042.
ALCF, a BG/P facility, was used to generate the gauge configurations used in this report.

We use standard ratios,  \(
C_{\rm 3pt}^{\Gamma,O}(t_{\rm sink}, t)/C_{\rm 2pt}(t_{\rm sink}),
\)
of the nucleon three-point, \(
C_{\rm 3pt}^{\Gamma, O}(t_{\rm sink}, t) =
\sum_{\alpha, \beta}
\Gamma_{\alpha\beta}
\langle
N_\beta(t_{\rm sink})O(t)\bar{N}_\alpha(0)
\rangle,
\)
to two-point, \(
C_{\rm 2pt}(t_{\rm sink}) =
\sum_{\alpha, \beta}
(1/2)(1+\gamma_t)_{\alpha\beta}
\langle
N_\beta(t_{\rm sink})\bar{N}_\alpha(0)
\rangle,
\)
functions to calculate the nucleon observables, \(O\), \cite{Lin:2008uz}.
We use \(N=\epsilon_{abc}(u_a^T C \gamma_5 d_b) u_c\) as our nucleon operator, Gaussian smearing \cite{Alexandrou:1992ti,Berruto:2005hg} for the source and point sink, and appropriate projections \(\Gamma\).
As was reported earlier, we optimized our gauge-invariant Gaussian-smearing width and source-sink separation, \(t_{\rm sink}\), so as to remove the excited-state contamination below our statistical error:
\begin{figure}[tb]
\includegraphics[width=.49\textwidth]{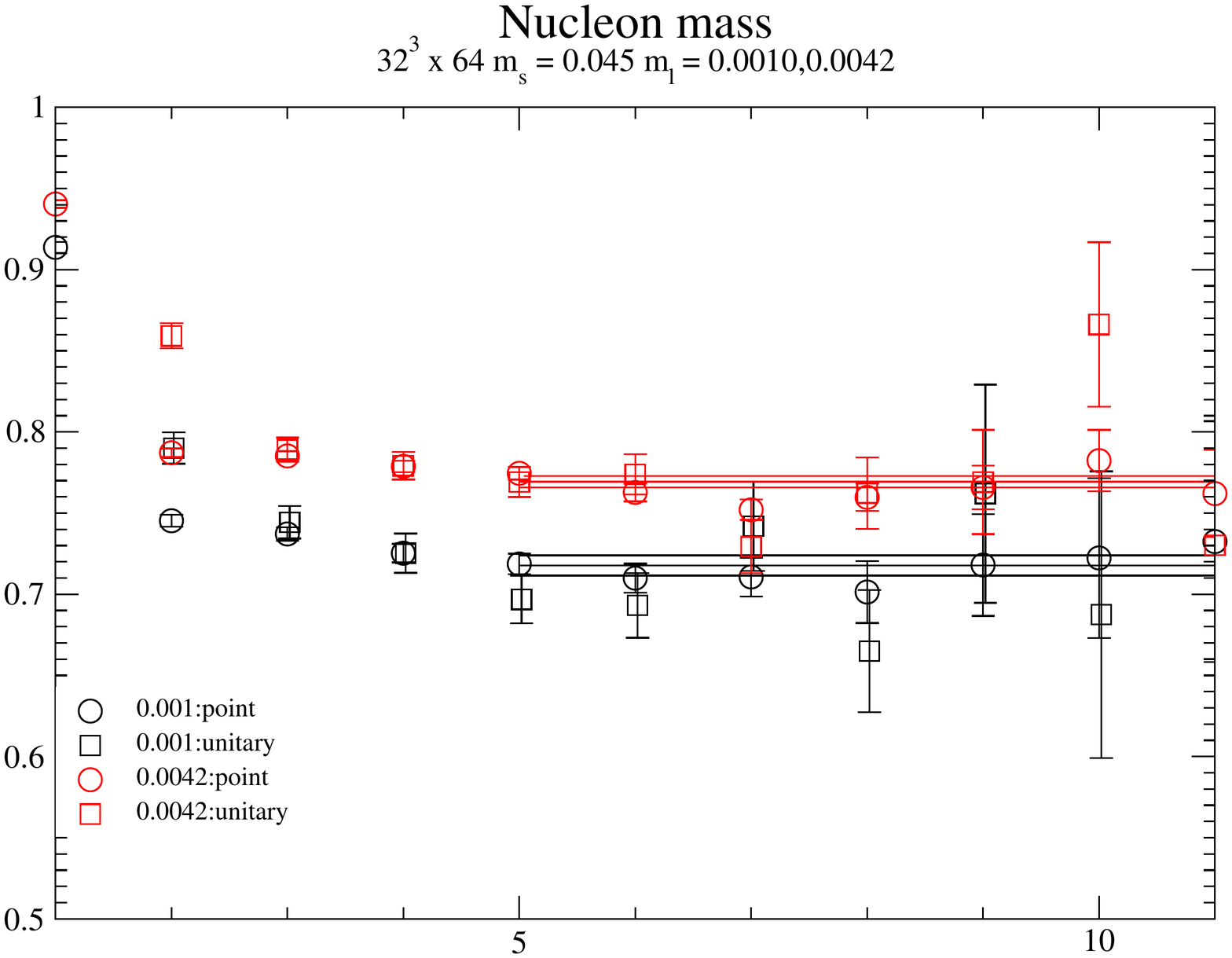}
\includegraphics[width=.49\textwidth]{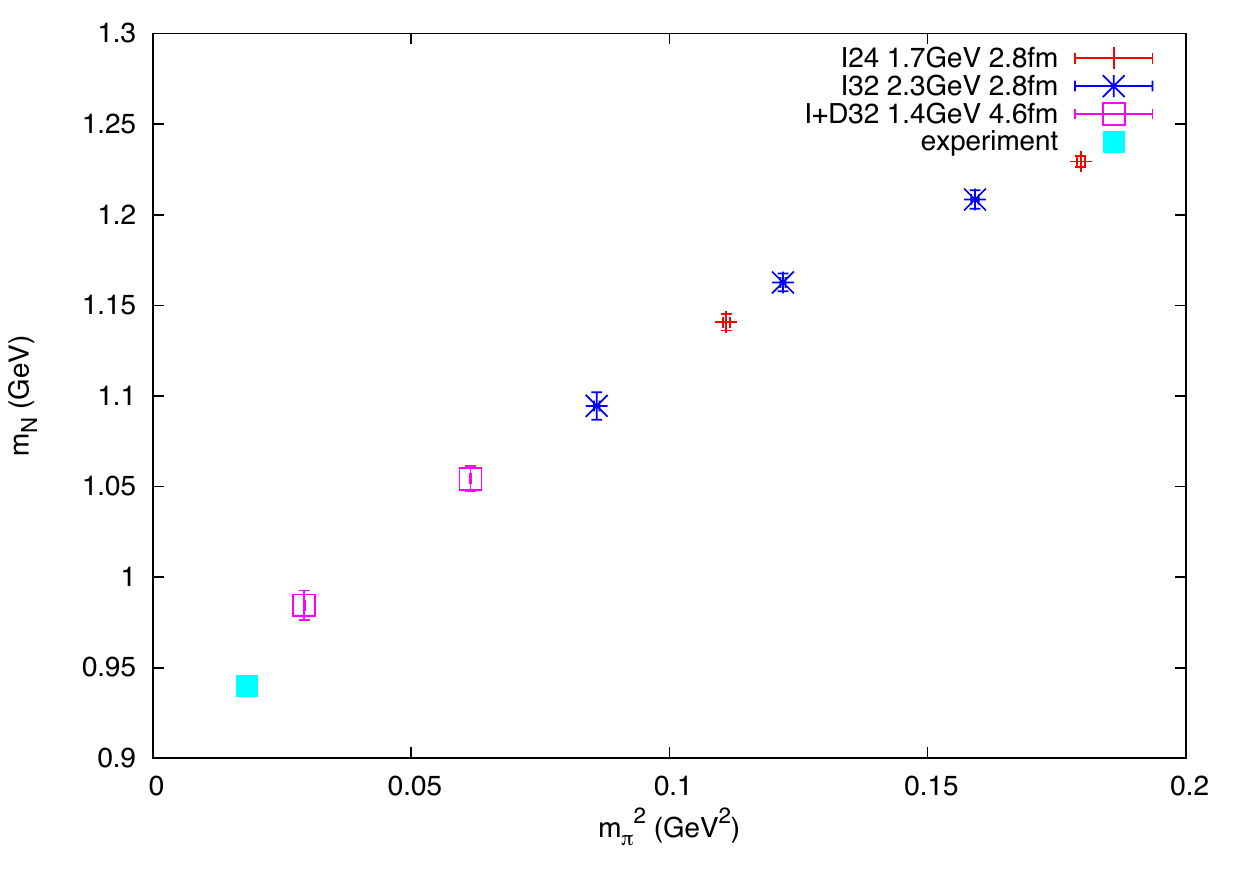}
\caption{
Left: nucleon mass effective-mass plot for RBC/UKQCD (2+1)-flavor dynamical ensembles with lattice cut off of \(a^{-1}\) = 1.371(8) GeV and pion mass \(m_\pi \sim 250\) MeV (red) and 170 MeV (black).  We estimate the nucleon mass as \(m_N=0.769(5)\) or 1.054(7) GeV for the former and \(m_N=0.718(6)\) or 0.984(8) GeV for the latter.  Right: \(m_N\) plotted against \(m_\pi^2\) from recent RBC+UKQCD 2+1-flavor dynamical DWF ensembles.
}
\label{fig:mNsignals}
\end{figure}
we chose the Gaussian width of six, and sour-sink separation of nine \cite{Ohta:2010sr,Ohta:2011vv,Lin:2011vx}.
We use 165 configurations for the heavy, \(m_\pi = 250\) MeV, ensemble, every eighth trajectory from the trajectory 608 to 1920, and 114 configurations for the light, 170-MeV, ensemble, from 508 to 1412.
For this report we have four source positions at \(t=0\), 16, 32 and 48 for each configuration.
The calculations are done on the RIKEN RICC computer cluster for the former, and NSF Teragrid/XSEDE clusters for the latter.
We obtain nucleon mass estimates of \(m_N=0.769(5)\) or 1.054(7) GeV for the former and \(m_N=0.718(6)\) or 0.984(8) GeV for the latter (see Fig.\ \ref{fig:mNsignals}.)
These values fall on a smooth line with our earlier results to the physical point.

\section{Vector and axial charges and their ratio}

\begin{figure}[t]
\includegraphics[width=.49\textwidth]{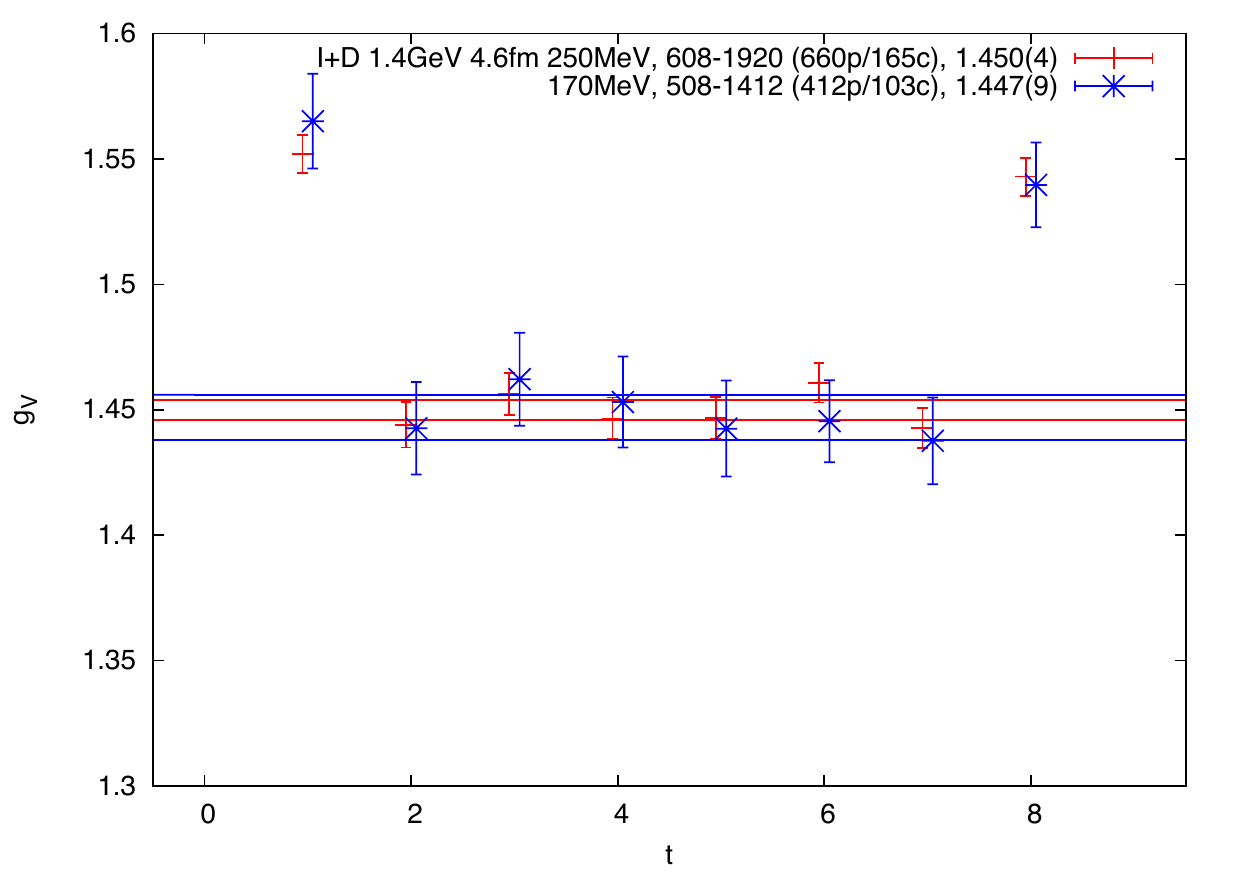}
\includegraphics[width=.49\textwidth]{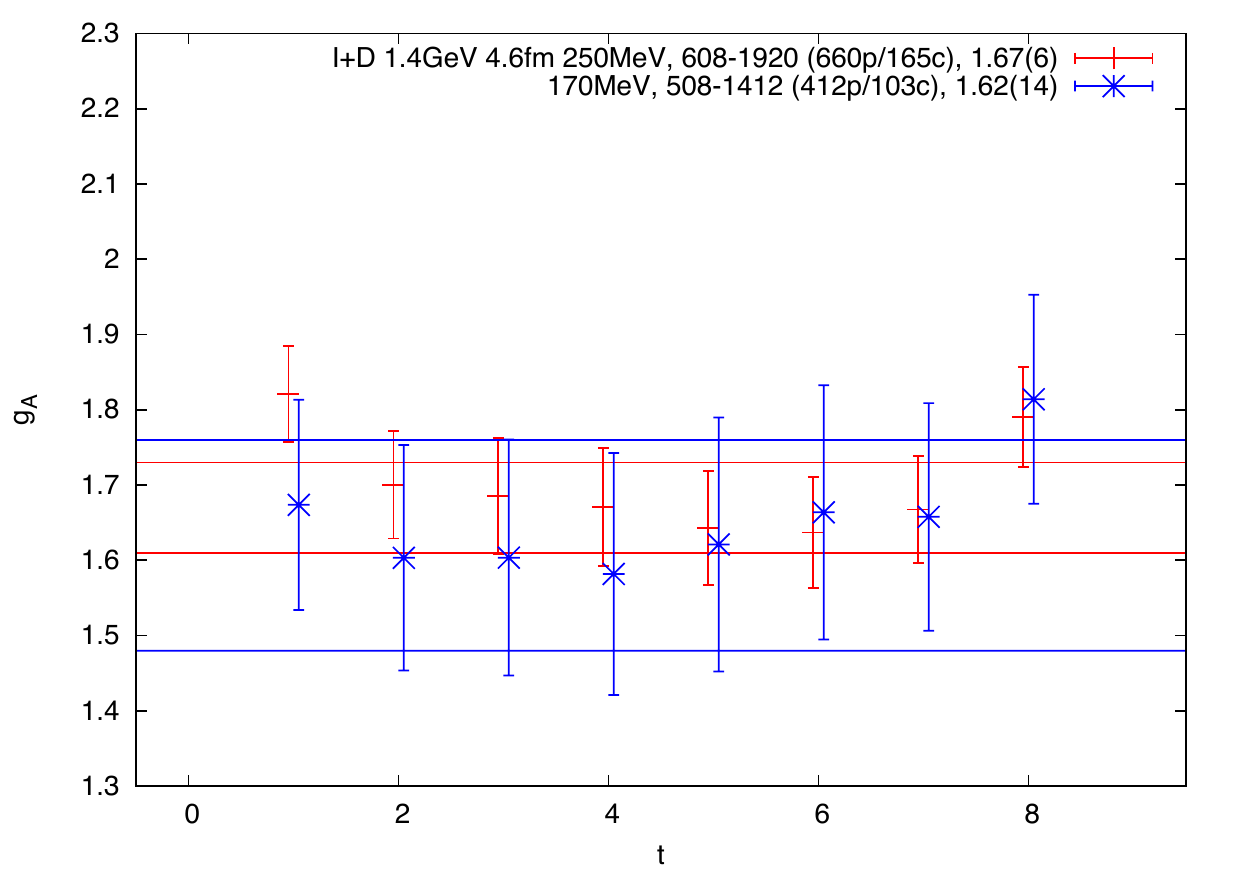}
\caption{
Left: local-current isovector vector charge gives very clean and accurate signals.
We obtain estimates of \(g_V = 1.450(4)\)  or 1.447(9), corresponding to \(Z_V = 0.692(7)\) in the chiral limit, in good agreement with independent estimates of  \(Z_V = 0.673(8)\) and \(Z_A = 0.6878(3)\) obtained in the meson sector.
This demonstrates  good chiral and flavor symmetries, up to small \(O(a^2)\)  correction, of the present ensembles even at relatively low cutoff of 1.371 GeV.
Right:local-current isovector axial charge.
}
\label{fig:chargesignals}
\end{figure}

We calculate the isovector vector charge, \(g_V\), of nucleon from the time component of the local current.
It deviates from unity and gives the inverse of the non-perturbative renormalization, \(Z_V\), of the current: \(g_V = 1/Z_V\), up to small \(O(a^2)\) correction.
As is shown in the left pane of Fig.\ \ref{fig:chargesignals}, we obtain very clean and accurate determination of \(g_V = 1.450(4)\) for the heavy, 250-MeV, ensemble and 1.447(9) for the light, 170-MeV, one, that extrapolate to a value \(g_V^{-1} = 0.692(7)\) in the chiral limit, in good agreements with  values obtained independently in the meson sector,  \(Z_V = 0.673(8)\) and \(Z_A = 0.6878(3)\) \cite{Arthur:2012yc}.
This demonstrates the good chiral and flavor symmetries of these ensembles at this relatively low momentum cut off of \(a^{-1} = 1.371(8)\) GeV.

The isovector axial charge, \(g_A\), is also calculated by the corresponding local current, this time from averaging its space components.
The results, shown in Fig.\ \ref{fig:chargesignals} in the right pane, are noisier than the vector charge, especially for the light ensemble which needs more statistics.
Fortunately the signal for the heavy ensemble is sufficiently good to draw important conclusions as we discuss in the following.

\begin{figure}[tb]
\begin{center}
\includegraphics[width=.7\textwidth]{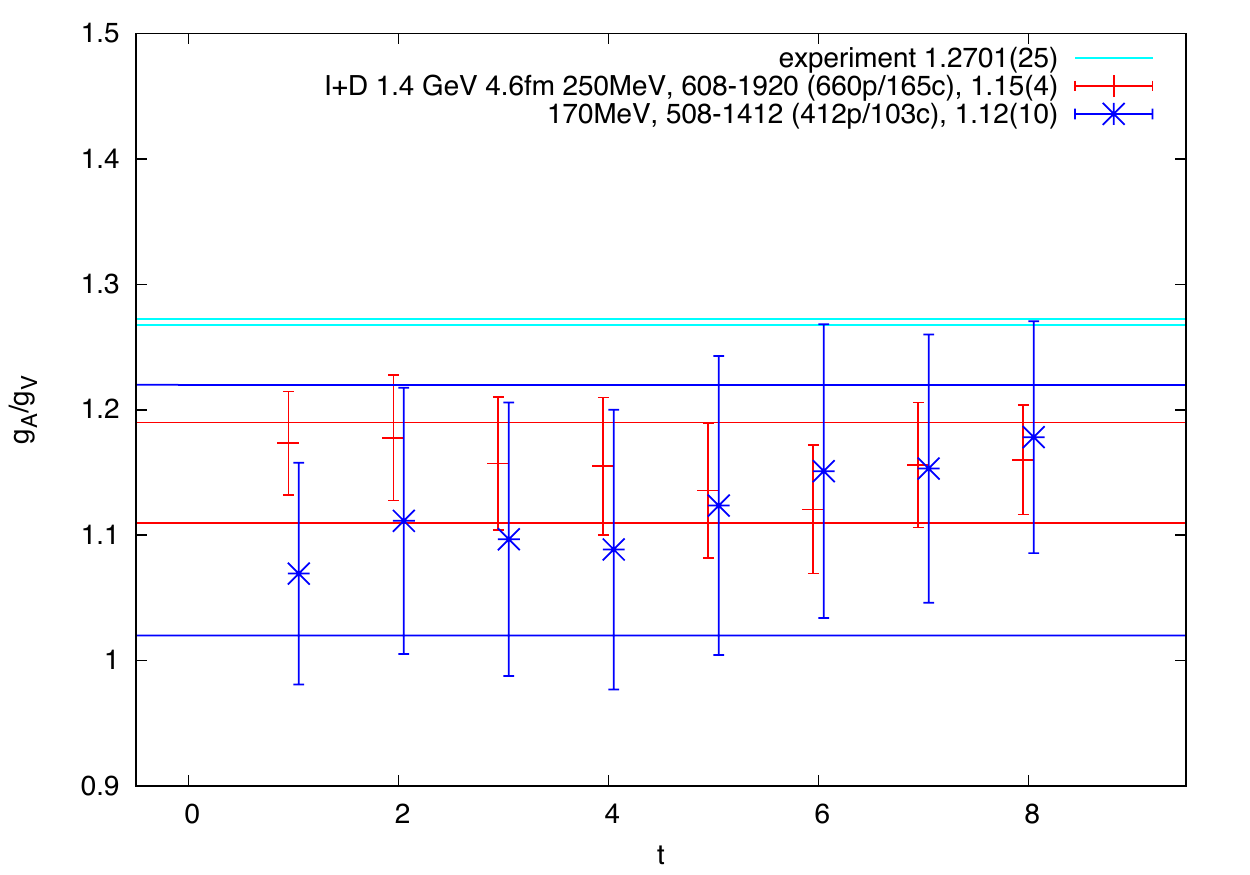}
\end{center}
\caption{
Ratio, \(g_A/g_V\), of the isovector local-current axial and vector charges.
Good chiral and flavor symmetries of DWF formulation make this ratio naturally renormalized and provide clean and accurate signals.  We obtain estimates of 1.15(5) and 1.12(10).
}
\label{fig:gAgVsignals}
\end{figure}

The ratio, \(g_A/g_V\), of isovector axial and vector charges, shown in Fig.\ \ref{fig:gAgVsignals}, is much less noisy in comparison with the axial charge itself.
We obtain estimates of \(g_A/g_V = 1.15(5)\) for the heavy, 250-MeV, and 1.12(10) for the light, 170-MeV, ensembles, respectively.
As this ratio is naturally renormalized under the good chiral and flavor symmetries of the DWF formalism, up to small \(O(a^2)\) correction, these values can be compared directly with the experiment of 1.2701(25) \cite{PDG2012}: they are about 10-\% below the experiment.  The deficit is almost three-standard-deviation significant for the former, heavy ensemble, though the value for the light ensemble needs more statistics to be significant.

\begin{figure}[tb]
\begin{center}
\includegraphics[width=.7\textwidth]{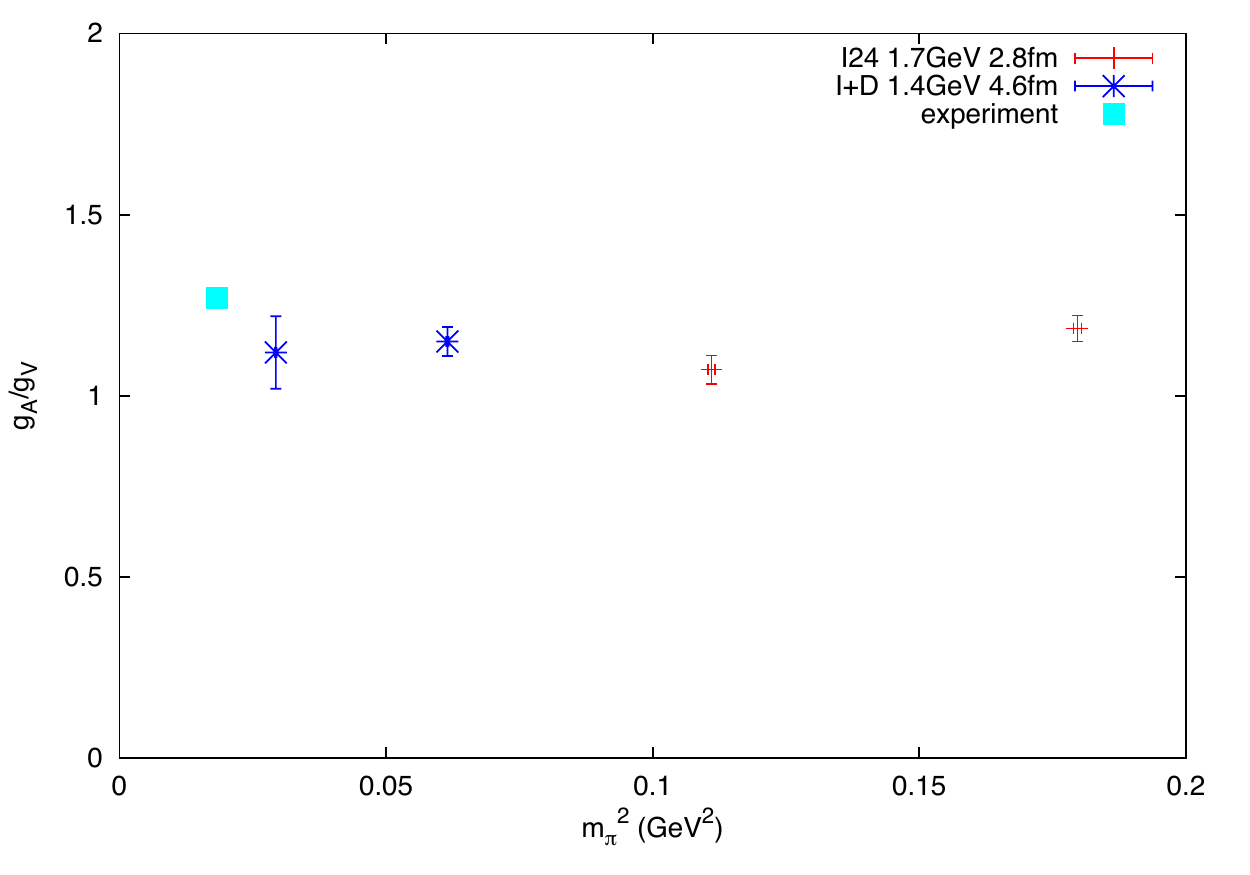}
\end{center}
\caption{
Non-monotonic dependence of ratio, \(g_A/g_V\), on \(m_\pi^2\): it once decreases with \(m_\pi^2\) in 1.7-GeV ensembles (red) but returns to an indistinguishable value at a lighter mass in 1.4 GeV ensemble (blue).
}
\label{fig:AVmpi2}
\end{figure}

Excited-state contamination has been proposed as a possible explanation for this discrepancy \cite{Wittig:2012np,Capitani:2012gj}.
The excited-state spectrum depends rather strongly on pion mass:
As the separation between the ground state and the first excited state is expected to grow with decreasing pion mass, such contamination would decrease.
Note also the way we calculate the vector charge, \(g_V\), protect it from the excited-state contamination: conserved charge cannot excite the ground state.
In Fig.\ \ref{fig:AVmpi2} we plot our values of the ratio,  \(g_A/g_V\), against the pion mass squared.
It is not monotonic:
If we compare the two points from our previous calculations with Iwasaki action, presented in red in the figure, the deficit seems to grow with decreasing pion mass.
But the heavier of the two present ensembles, presented in blue, brings the value back to less deficit, to a value the difference of which is statistically insignificant.
This non-monotonic behavior is hard to explain if the significant excited-state contamination were present in either of these ensembles, as such contaminations have to agree across the ensembles generated with different gauge actions and physics parameters.

The picture becomes much simpler when we plot the ratio, \(g_A/g_V\), against the finite-size scaling parameter, \(m_\pi L\), as is shown in Fig.\ \ref{fig:gAgVmpiL}:
\begin{figure}
\begin{center}
\includegraphics[width=.7\textwidth]{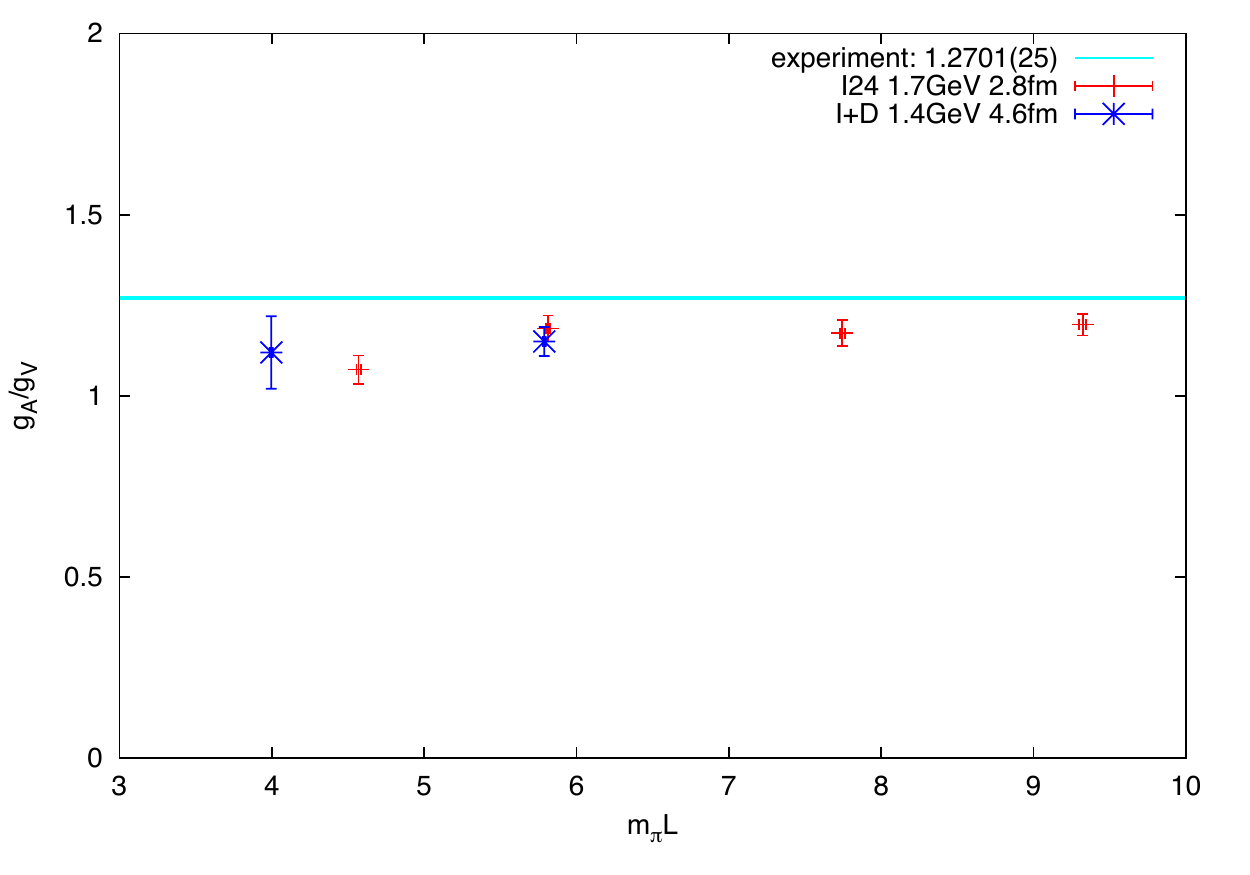}\\
\end{center}
\label{fig:gAgVmpiL}
\caption{
Ratio, \(g_A/g_V\), of the isovector axial and vector charges, plotted against the dimensionless finite-size scaling parameter, \(m_\pi L\).
THe dependence is monotonic, and two results at similar values of \(m_\pi L \sim\) 5.8, 1.19(4) from I24 and 1.15(5) from ID, agree with each other, despite very much different \(m_\pi\) and lattice cuts-off, that significantly alter mass spectrum.
}
\end{figure}
All the results from our earlier calculations with Iwasaki gauge action at cut off momentum of about 1.7 GeV and the present ones seem to align well monotonically in this plot, with the deficit growing with decreasing \(m_\pi L\).
And more importantly the two points that share about the same value of \(m_\pi L\sim 5.8\), one from the earlier Iwasaki calculation with \(m_\pi = 420\) MeV and \(L=2.8\) fm, and another from the present calculation with \(m_\pi = 250\) MeV and \(L=4.6\) fm, agrees within each other's statistical errors, 1.19(4) from the former and 1.15(5) from the latter, despite their quite different gauge action, cut off momenta, DWF residual mass, source-sink separations, quark mass, and spatial volumes.
They are very unlikely to agree if one or the other or both suffered significant excited-state contamination, as such contamination would have to agree with high accuracy despite their quite different gauge actions, cut off momenta, DWF residual mass, source-sink separations, quark mass, and spatial volumes.

\section{Conclusion}

RBC and UKQCD collaboration continue to calculate nucleon-structure observables using the 2+1f dynamical DWF ensembles they jointly generate.
We found the following from the ongoing calculations with ensembles with latticee cut off momentum of \(a^{-1} = 1.371(8)\) GeV, the spatial extent \(L = 4.6\) fm and pion mass \(m_\pi = 250\) and 170 MeV:
\begin{enumerate}
\item Nucleon mass estimate is \(m_N \sim 1.054(7)\) and 0.984(8) GeV respectively for the heavy and light ensembles.
\item Isovector vector charge is well-behaved, demonstrating good chiral and flavor symmetries with DWF at this relatively low cut off momentum.
\item Isovector axial charge is noisier, yet the calculation is solid.
\item There is about 10\% deficit in \(g_A/g_V\), with almost 3-standard-deviation significance.
\end{enumerate}
By comparing the result from the heavy ensemble with our earlier calculation with pion mass of about 420 MeV and lattice extent of about 2.8 fm that shares similar value of the finite-size scaling parameter, \(m_\pi L \sim 5.8\), we find excellent agreement in \(g_A/g_V\): 1.15(5) from the present and 1.19(4) from the earlier, despite their coming from different gauge actions, lattice cut off momenta, source-sink separations, pion mass and lattice volume.
Such an agreement is unlikely if there were significant excited-state contamination present in either of the calculations.
Thus we conclude neither suffers from such a contamination above our statistical error.

This confirms the scaling in finite-size parameter, \(m_\pi L\).
The result can be interpreted as the first concrete evidence for the pion cloud surrounding nucleon.
With light pion cloud surrounding it, nucleon is hardly point-like: at \(m_\pi=250\) MeV the 4.6 fm linear extent of the present lattice is insufficient to contain the axial current carried by the cloud.
It grows further toward the lighter physical mass of  \(m_\pi \sim 140\) MeV.
To sufficiently contain the axial charge, spatial lattice extent much larger than the 4.6 fm is required.
This seems hard to reconcile with the conventional nuclear models with point-like nucleons interacting through non-relativistic potential.

In the talk the status of our structure function study was briefly summarized as well:  signals for the low moments of structure functions such as isovector quark momentum and helicity fractions are noisier than those for form factors, but calculations for these are well under way.

We are increasing our statistics: double at least by adding extra source coordinates at \(t=8\), 24, 40, and 56, and possibly more if necessary, for the heavy ensemble.  
We plan at least the corresponding size of statistics for the light ensemble, using the new AMA technique \cite{Eigo:Cairns2012}.  
We seek calculations at physical pion mass, and with appropriate isospin breaking soon afterward.

\end{document}